# A possibility of increase of a superconducting gap in a two-dimensional BCS superconductor

I. N. Zhilyaev
*Institute of Microelectronics Technology and High Purity Materials, Russian Academy of Sciences, Chernogolovka, Moscow Region, Russia*
zhilyaev@ipmt-hpm.ac.ru

According to the BCS model the probability of filling of pair electron states $v_k^2$ as a function of kinetic energy is smeared i.e. is not a step function. This means that part of energy of connection through virtual phonons is consumed to increase the kinetic energy in an electronic system at the transition to the superconducting state. In this work a modified mechanism of $v_k^2$ formation is proposed which takes into account the dynamics of fluctuations near the phase transition. The estimations of a superconducting gap and isotopic effect for a material with parameters of high-temperature superconductors give values comparable with these observed.

In spite of a great number of proposed explanations of high-temperature superconductors (HTS) nature, no generally accepted mechanism has yet been formulated. Experimental data show that high superconducting properties of HTS are retained down to one layer [1] and, consequently, the state of HTS electronic system can be considered two-dimensional in the first approximation. In fact situation is more complicated because the HTS electronic structure in each layer is non-uniform [2] and is strongly anisotropic. These factors complicate the construction of an exact theory. Moreover, the consideration of fluctuation influence in such a conductor necessarily involves the account of the effect of fluctuation dynamics which cannot be realized now [3]. Therefore, phenomenological models of a superconducting state of a two-dimensional conductor with HTS properties where the dynamics of fluctuations is taken into account seem justified. Such a model is proposed in the work.

Cooper suggested [4] a state of two electrons connected by virtual phonons (cooper pairs, "pairs" in the text to follow), existing in a metal above the Fermi level. According to the BCS model [5,6], in a superconducting ground state electron pairs interact through virtual phonons in the Debye energy range $\hbar\omega_D$ near the Fermi level. For such processes to occur, the interval should involve free states. As a result, the probability of filling of pair electron states $v_k^2$ as a function of electron energy $\varepsilon$ does not break abruptly at a level corresponding to the Fermi energy $E_F$ even at temperature $T = 0$, as it occurs in a normal metal, but is smeared by kinetic energy. Numerically, energy smearing will be regarded not as a $\hbar\omega_D$ magnitude, but rather as an energy interval $\delta\varepsilon$ near the Fermi level in which the function $v_k^2$ does not strongly differ from the value 1/2. The fact is that in superconductors described by the BCS theory, $\delta\varepsilon = \delta\varepsilon_{wc} \ll \hbar\omega_D$ in the case of weak connection (wc), which will be of interest for in this work. In calculating a superconducting gap $\Delta$, the basic contribution to the sum and final value $\Delta$ are due to the interval $\delta\varepsilon_{wc}$ although the states summation is in an energy interval of the order $\hbar\omega_D$. This contribution of $\delta\varepsilon_{wc}$ is connected with a relatively high value of the product $v_k^2(1-v_k^2)$. The physical sense of this is that processes between electrons are most probably to occur in this interval, resulting in the formation of a superconducting state. According to the BCS model $\Delta$ increases with increasing $\delta\varepsilon_{wc}$ (for example, when the electron interaction potential V or density of states $n(E_F)$ at the Fermi level increase). The idea of the BCS model modification proposed here consists in considerable increase of $\delta\varepsilon$ as compared with the initial interval $\delta\varepsilon_{wc}$ rather than the electron interaction potential V and density of states, in the case of weak connection. To put it another way, to increase an energy interval where the product $v_k^2(1-v_k^2)$ is rather great.

Let first estimate the maximum value $\delta\varepsilon$ at which the formation of a superconducting state is possible. Then admit that for some reason there occurs a pair smearing described by the function $v_k^2$ at a given moment in a large interval $\delta\varepsilon > \delta\varepsilon_{wc}$. In principle, a superconducting state can be formed by this smearing. But the state described by the function $v_k^2$ is non-stationary. In the frame of the theory of Fermi liquid, an electron lifetime at a level $\delta\varepsilon/2$ away from the Fermi level is $\tau_\varepsilon \cong 2\pi\hbar E_F/(\delta\varepsilon/2)^2$ (see for example [7]). With time, the current distribution $v_k^2$ would transform so that smearing $\delta\varepsilon$ would decrease and transform into smearing $\delta\varepsilon_{wc}$, corresponding to the BCS model. The minimum lifetime corresponds to the state most distant from the Fermi level as follows from the formula for $\tau_\varepsilon$. The formation of a superconducting state also proceeds at certain characteristic times. Pairing interaction between electrons occurs by means of virtual phonons which are characterized by the Debye frequency $\omega_D$. Let consider that the lifetime does not prevent the formation of a superconducting state when it is five times larger than the inverse Debye frequency, $\tau_\varepsilon \geq 10\pi/\omega_D$. The maximum value of smearing is then estimated from the equation $\tau_\varepsilon = 10\pi/\omega_D$. Taking $E_F$ equal to 0.3 eV typical for HTS, we have $\delta\varepsilon/2 = 700K$. From the formula of the BCS model the gap value can be estimated as

$$\Delta = V\sum_k v_k\sqrt{(1-v_k^2)}$$

Consider then the function $v_k^2$ given and equal to 1/2 in the smearing limits. Neglecting anisotropy and replacing states summation by energy $\varepsilon$ integration, we have in the smearing limits:

$$\Delta = V \int n(E_F)\sqrt{1/2}\sqrt{(1-(\sqrt{(1/2)})^2)}d\varepsilon$$

Taking $v_k^2 = 0$ beyond the smearing interval from $\varepsilon_1 = -700K$ to $\varepsilon_2 = 700K$, we have

$$\Delta = V \int_{\varepsilon_1}^{\varepsilon_2} n(E_F)(1/2)d\varepsilon = Vn(E_F)(1/2)1400K$$

At $Vn(E_F) = 1/2$ for a case of weak link [8] we obtain $\Delta \cong 350$ K. Thus, large smearing, even in the case of weak link, can give a gap value comparable with the value observed in HTS.

To maintain large non-stationary smearing, use can be made of fluctuations of charge density waves (CDW) or spin density waves near the phase transition. This idea can be explained on the example of CDW for a conductor in a normal state. In [9] arguments are adduced for the presence of a quantum critical point (QCP) in HTS. Let take QCP in our conductor present, connected with the phase transition to a CDW state. CDW state is due to the potential caused by the exchange and correlation interaction in an electronic system [10]. CDW fluctuations occur near QCP. During fluctuation a potential and a redistribution of a charge occur in a place of its positioning, i.e. electric currents arise. As a result, the kinetic energy in an electron system changes. Electrons, running through the location of fluctuation and remaining unconnected in the CDW state, would change the kinetic energy by a certain value. At a free length path, greater than the size of fluctuation, the kinetic electron energy, owing to fluctuation, would be driven beyond the limits of fluctuation for a distance about the order of the free length path. This would lead to the smearing of kinetic energy of electrons residing outside the fluctuation. In greater detail, the essence of smearing formation can be considered on an example of a relatively simple electronic structure with a symmetry of the second order. So, assume, that no interaction between electrons through virtual phonons, which would transform a normal state into superconducting. Let QCP as phase transition to the CDW state arise in the considered two-dimensional conductor at $T = 0$ and reduction of charge carriers concentration. At a concentration higher than QCP the conductor has a Fermi surface (or the contour, to be more precise) in the form close to a rectangular. The electronic spectrum of the conductor is transformed so that electrons form CDW as a spatial charge modulation with the period d at lowering of the concentration of charge carriers. Then the spectrum of the conductor should have a CDW gap which size smoothly varies depending on the direction in pulse space, falling down from the maximum value $\varepsilon_{cdw}$ in the direction of spatial charge modulation to zero in the perpendicular direction.

Heterogeneity in the form of strips 20÷30 Å wide [2], probably connected with CDW are observed in HTS by atomic force microscopy. Below we shall consider d = 25Å. An anisotropic pseudo-gap can also exist in HTS, which maximum size in the range of optimum doping can be about 1000K [11]. Let consider $\varepsilon_{cdw}$ in our case be of the same order of magnitude. To estimate smearing we should know how electron energy can change at the interaction with CDW fluctuation. To this end, let consider the picture of fluctuation. At temperatures close to zero and charge concentrations corresponding to a value before the transition to a CDW state, CDW fluctuations in a conductor should occur as spatial redistributions of a charge in the direction of CDW charge modulation. For the sake of simplification in this case we have, first, chosen low temperatures to avoid accounting the presence of real phonons. Second, T is chosen ≠ 0 not to take into account the specific character of quantum fluctuations near QCP. Let fluctuations having the minimum scale about d/2 ÷ d are present under these conditions (figure). During fluctuations, spatial redistribution of a charge occurs in two systems: in an electron system and in an ionic lattice which tends to compensate for non-uniform distribution of a charge in the electron system [10]. Electron and ion masses differ by several orders of magnitude. Therefore, although the process of CDW fluctuations occurs due to the potential in an electron system, the site of fluctuations is determined by a more inert ionic lattice. To estimate characteristic times and energy, let describe fluctuations in electron and ionic systems as processes of oscillations. For relatively fast fluctuations in electron system, spatial fluctuations of the lattice are actually static. That is, electronic fluctuations should be located in the space on a quasistatic relief of spatial modulation of a d/2 ÷ d lattice charge (figure). The coulomb interaction arises during spatial redistribution of electron density which should result in the plasma law of fluctuations [12]. Let consider processes of fluctuations in an electron system as plasma waves. Plasma waves in a two-dimensional conductor are described by the law of dispersion $1/\omega^2 = 1/(c/\lambda)^2 + 2/\omega_p^2$ [13], where c - velocity of light, $\omega_p = (4\pi n_e e^2/m)^{1/2}$ - plasma frequency of a three-dimensional conductor (e and m are the charge and weight of an electron). The estimation of plasma frequency of charge density in HTS near optimum doping $n_e = 5*10^{21}$ см$^{-3}$ gives $\omega_p \cong 4*10^{15}$ radian/sec. The wave length λ cannot be larger than the size of fluctuation. It then follows from the law of dispersion that fluctuations can have only one frequency $\omega_s = \omega_p/\sqrt{2} = 2.8*10^{15}$ radian/sec. Such fluctuations can exist only at small energy losses in an electron system. Let consider possible channels of relaxation in an electron system and conditions when they are small. In a conductor processes of scattering of charge carriers can occur in the volume and on the surface. The free length path is usually considered to be about 100Å even in a "dirty" limit. The effect of volume scattering on impurities and defects on oscillations is small at $\lambda < 25\text{Å} < 100\text{Å}$. At the diffusive character of charge carrier scattering on the surface, it can influence the components of electron impulses. Cross electronic focusing [14] revealed mirror scattering of holes in bismuth. Let charge carriers of appropriate sign undergo mirror scattering in our case, that is, losses on the surface can be neglected. To estimate the kinetic energy of fluctuations in an electron system under the conditions of low attenuation, let consider the fluctuations as quantum oscillator. Then the energy of the order $\hbar\omega_s$ corresponds to frequency $\omega_s$. For estimation, take the kinetic energy $E_{kin}$ describing fluctuation be an oscillator of the magnitude $\hbar\omega_s$. This energy, in turn, corresponds to a certain number N of electrons. Now, estimate the energy $\varepsilon_{kin}$, per one electron. First, estimate the number N. Taking the thickness of a conducting layer equal 12Å - typical value of the lattice parameter across layers in HTS, and N as the number of electrons in the volume $D = (25/2)\ 25\cdot12\text{Å}^3$, we obtain $N \cong 20$. Then $\varepsilon_{kin}$ is equal to $\hbar\omega_s/N \cong 1000K$. The fact that the values of $\varepsilon_{kin}$ and $\varepsilon_{cdw}$, proportional to kinetic and potential energy, are close suggests that the described fluctuations are quite possible in an electron system. At the Fermi velocity $v_F \cong 10^7$ cm/sec typical in HTS the time of electron interaction with fluctuation $d/2v_F$ is appreciably larger than the period $2\pi/\omega_s$ describing oscillations. In this situation, fluctuations in an electron system can transfer running electrons to an energy state either above or lower than their previous state with an arising electric field to states which are not populated. Energy transferring is mainly upwards because lower energy states are more densely populated. Thus, smearing is formed in an electron system. Let now estimate the magnitude of smearing which can be supported by CDW fluctuations. Assume some point in the phase diagram where the time of electron free path from fluctuation to fluctuation is $\tau_e \cong 10\pi/\omega_D$. Take energy relaxation into account [7]. Now compare the balance of energy pumping and relaxation of running electrons between fluctuations. Certainly, not all the energy of fluctuation can be transferred to escaping electrons. For estimation, take the energy of each electron at the edge of smearing where smearing proceeds during the interaction with fluctuations increases by $\delta\varepsilon_{kin} = \varepsilon_{kin}/2 = 500K$ within the time $\tau_e$. Let the rate of energy relaxation under the conditions of nonstationary electron states [7] be determined by the lifetime $\tau_n \cong 2\pi\hbar E_F/(\delta\varepsilon_n/2)^2$ corresponding to the smearing edge, where $\delta\varepsilon_n/2$ is the difference between the state energy at the smearing edge and Fermi level. Assume that the rates of energy increase and energy relaxation are on average equal in in time $\delta\varepsilon_{kin}/\tau_e \cong (\delta\varepsilon_n/2)/\tau_n$ for electrons running between fluctuations at smearing edge. It follows that $\delta\varepsilon_n/2 = (\delta\varepsilon_{kin}\omega_D 2\pi\hbar E_F/10\pi)^{1/3}$. Then for energy smearing is $\delta\varepsilon_n/2 \cong 620K$. If such smearing were maintained in a superconducting state ($\delta\varepsilon = \delta\varepsilon_n$), the gap would be $\Delta = 310K$. But in a superconducting state at $\delta\varepsilon_{kin} = 500K < 2\Delta = 620K$ electrons cannot derive energy from fluctuations to support smearing because the energy $\delta\varepsilon_{kin}$ is too small to disrupt the bonding in a pair. Therefore the value $2\Delta$ cannot be larger than $\delta\varepsilon_{kin}$. When $2\Delta < \delta\varepsilon_{kin}$ the interaction between part of electrons and fluctuations can proceed via the disruption of bonding in pairs and increase of electron energy. This would result in an increase of smearing and $2\Delta$ to the level $\delta\varepsilon_{kin}$ because the rate of superconducting state formation is higher than the rate of electron interaction with fluctuations. As a result the only possible variant in this case is $\Delta \cong \delta\varepsilon_{kin}/2 = 250K$. Because the maximum value of $\Delta$ determines $\delta\varepsilon_{kin}$, the dependence $\Delta(n_e)$ should exhibit a maximum not as point but as a plateau with $\Delta \cong \delta\varepsilon_{kin}/2 = 250K$, when the charge carrier concentration changes.

Some parallels seem relevant between the properties of the proposed superconducting state and the properties characteristic of HTS but qualitatively different from the properties conventional BCS superconductors. These are:

1) In HTS a pseudo-gap is observed. In the proposed model to it corresponds energy smearing for the probability of filling of electronic states, connected with CDW fluctuations.

2) In HTS, $\Delta$ drops as doping increases or decreases from an optimum level. In the proposed model such a behavior can be caused by an increased distance from the phase transition. Also note that the region of quantum phase transition can be very wide [15], which are possible in HTS as well.

3) In HTS with high superconducting parameters in the region of optimum doping the isotopic effect is close to zero [16]. Under the conditions of the proposed model the isotopic effect is equal to zero.

4) In HTS Bi-2212 the dependence of $\Delta$ on the hole concentration exhibits a slight flattening in the region of optimum doping [17]. According to the proposed model the maximum of $\Delta(n_e)$ dependence should be a plateau.

5) The proposed model can work in the case of the fourth order symmetry when the displacement of charges at CDW fluctuations occurs in two perpendicular directions.

6) The proposed model can probably be applied to the case of spin density wave. Electromotive forces should arise at fluctuations of magnetization in a conductor. The last, in turn, can bring about smearing.

This work was supported by the Programs "Quantum physics of condensed media" and "Computing by new physical principles".
The author is grateful to V.A.Tulin, V.I.Yudson, V.T.Dolgopolov, V.V.Ryazanov, V.Ya.Kravchenko, V.B.Shikin, V.M.Edelshtein, A.V.Haetskiy, L.N.Zherihina, S.A.Brazovskiy, K.B.Efetov, I.A.Larkin, V.G.Popov, Yu.V. Kopaev, Yu.I.Latishev, S.V.Zaitsev-Zotov, M.Yu.Barabanenkov for the consultations and useful discussion.

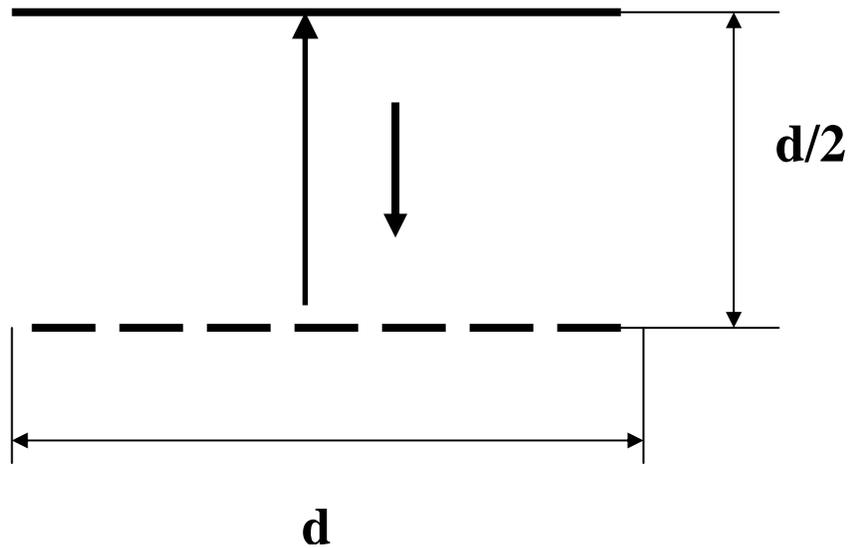

Figure caption: the schematic image of CDW fluctuation. Lines in the figure are CDW extremum. The arrows show the directions of currents of free charges and ions at the formation of CDW fluctuation.